\documentclass[aps,prd,twocolumns,eqsecnum,preprintnumbers,showpacs,amsmath,amssymb]
{revtex4}

\usepackage{graphicx}
\usepackage{dcolumn}
\usepackage{bm}

\begin{document}

\title{I. THE MASS GAP AND SOLUTION OF THE \\ QUARK CONFINEMENT PROBLEM IN QCD}

\author{V. Gogokhia}
\email[]{gogohia@rmki.kfki.hu}

\affiliation{HAS, CRIP, RMKI, Depart. Theor. Phys., Budapest 114,
P.O.B. 49, H-1525, Hungary}

\date{\today}
\begin{abstract}

Using the previously derived confining gluon propagator, the
corresponding system of equations determining the quark propagator
is derived. The system of equations consists of the
Schwinger-Dyson equation for the quark propagator itself, which
includes the zero momentum transfer quark-gluon vertex. It is
complemented by the Slavnov-Taylor identity for this vertex. The
quark equation depends explicitly on the mass gap, determining the
scale of the truly nonperturbative dynamics in the QCD ground
state. The obtained system of equations is manifestly
gauge-invariant, i.e., does not depend explicitly on the
gauge-fixing parameter. It is also free from all the types of the
perturbative contributions ("contaminations"), which may appear at
the fundamental quark-gluon level.

\end{abstract}

\pacs{ 11.15.Tk, 12.38.Lg}

\keywords{}

\maketitle

\section{Introduction}

To say today that Quantum Chromodynamics (QCD) is the
nonperturbative (NP) theory of quark-gluon interactions is almost
a tautology. The problem is how to define it exactly, since we
know for sure that QCD has a nontrivial perturbative (PT) phase as
well because of asymptotic freedom (AF) \cite{1}. The
corresponding asymptotic mass scale parameter $\Lambda_{QCD} =
\Lambda_{PT}$ is responsible for its PT dynamics (scale violation,
AF, etc.). On the other hand, if QCD itself is a confining theory
then a characteristic scale is very likely to exist. It should be
directly responsible for the large-scale structure of the true QCD
vacuum in the same way as $\Lambda_{QCD}$ is responsible for its
short-scale one. However, the Lagrangian of QCD does not contain
explicitly any of the mass scale parameters which could have a
physical meaning even after the corresponding renormalization
program is performed.

The only place where the regularized version of the mass scale
parameter (the mass gap in what follow, for simplicity) may appear
is the dynamical system of quantum equations of motion of QCD. It
is known as the Schwinger-Dyson (SD) equations. They should be
complemented by the corresponding Slavnov-Taylor (ST) identities,
which relate the different Green functions, entering the SD
equations, to each other \cite{1}. To solve this system means to
solve QCD itself and vice-versa, since it contains the full
dynamical information on QCD (and even more than that). Some
solutions of these equations reflect the real structure of a QCD
ground state, which is necessary to know in order to understand
such an important physical phenomena as color confinement,
spontaneous breakdown of chiral symmetry (SBCS) and many other NP
effects. There is a close intrinsic link between these phenomena
and the true structure of the QCD vacuum \cite{2,3,4,5}.

Contrary to Quantum Electrodynamics (QED), in QCD the Green's
functions are essentially modified from their free counterparts
due to the strong response of the highly complicated structure of
the true QCD vacuum.  Such a substantial modification can be
neglected in the simplest cases only: in the weak coupling limit
due to AF or for heavy quarks. In other words, it is not enough to
know the Lagrangian of the theory. In QCD it is also necessary and
important to know the true structure of its ground state. This
knowledge comes just from the investigation of the above-mentioned
system of the SD equations and ST identities. Although this system
of dynamical equations can be reproduced by an expansion around
the free field vacuum, the final equations make no reference to
the vacuum of the PT. They are sufficiently general and should be
treated beyond the PT, and thus serve as an adequate and effective
tool for the NP approach to QCD \cite{1}.

 Also, we need these solutions for the
Green's functions in order to calculate the physical observables
in QCD from first principles. One of the main roles in the
realization of this program belongs to the solution for the gluon
Green's function which describes their propagation in the QCD
vacuum. In the presence of a mass gap responsible for the true NP
QCD dynamics it has been exactly established in our previous work
\cite{2} (for a brief review see below).

The main purpose of this work is to derive the confining quark
propagator on the basis of this solution by using the
above-mentioned system of the corresponding SD equations and the
quark-gluon ST identity.

\section{The confining gluon propagator}

In our previous work \cite{2} in order to realize the
above-mentioned mass gap responsible for the true NP QCD dynamics,
we propose not to impose the transversality condition on the full
gluon self-energy, while preserving the color gauge invariance
condition for the full gluon propagator. Since due to color
confinement the gluon is not a physical state, none of physical
observables/processes in QCD will be directly affected by such a
temporary violation of color gauge invariance/symmetry (TVCGI/S).
In order to make the existence of a mass gap perfectly clear the
corresponding subtraction procedure has been introduced. All this
allowed us to establish the general structure of the full gluon
propagator in the presence of a mass gap as follows (Euclidean
signature here and everywhere below):

\begin{equation}
D_{\mu\nu}(q) = i \left\{ T_{\mu\nu}(q)d(q^2) + \xi L_{\mu\nu}(q)
\right\} {1 \over q^2 },
\end{equation}
where $\xi$ is the gauge-fixing parameter and
$T_{\mu\nu}(q)=\delta_{\mu\nu}-q_{\mu} q_{\nu} / q^2 =
\delta_{\mu\nu } - L_{\mu\nu}(q)$. Evidently, it satisfies the
color gauge invariance condition $q_{\mu}q_{\nu}D_{\mu\nu}(q) = i
\xi$ as mentioned above (the color group indices, which in this
case is simply reduced to the trivial $\delta$-function, have been
omitted). The full gluon form factor or equivalently the full
effective charge $d(q^2) = \alpha_s(q^2)$ is then

\begin{equation}
d(q^2) = {1 \over 1 + \Pi(q^2; D) + (\Delta^2(\lambda; D) / q^2)}.
\end{equation}
Here $\Pi(q^2; D)$ is the subtracted full gluon self-energy, while
$\Delta^2(\lambda; D)$ is the difference between the full gluon
self-energy and its subtracted counterpart. Obviously, it is
nothing but the sum of all possible quadratic divergences
parameterized as the mass gap and regulated by $\lambda$.
Rewriting Eq. (2.2) as the corresponding transcendental equation
for the effective charge, we were able to formulate and develop
its nonlinear iteration solution \cite{2}. Finally it made it
possible to exactly decompose the regularized full gluon
propagator (2.1) as the sum of the two principally different terms

\begin{equation}
D_{\mu\nu}(q; \Delta^2) = D^{INP}_{\mu\nu}(q; \Delta^2)+
D^{PT}_{\mu\nu}(q),
\end{equation}
where

\begin{equation}
D^{INP}_{\mu\nu}(q, \Delta^2) = i T_{\mu\nu}(q) {\Delta^2 \over
(q^2)^2} f(q^2),
\end{equation}
and the superscript "INP" means intrinsically NP, while $f(q^2)$
is determined by the corresponding Laurent expansion as follows:

\begin{equation}
f(q^2) = \sum_{k=0}^{\infty} (\Delta^2 / q^2)^k \Phi_k(\lambda,
\alpha, \xi, g^2).
\end{equation}
The mass gap $\Delta^2 \equiv \Delta^2(\lambda, \alpha, \xi, g^2)$
depends on the same set of parameters as the residues
$\Phi_k(\lambda, \alpha, \xi, g^2)$ in the Laurent expansion
(2.5), where in addition $\alpha$ and $g^2$ are the dimensionless
subtraction point and the coupling constant squared, respectively.

The PT gluon propagator

\begin{equation}
D^{PT}_{\mu\nu}(q) = i \Bigr[ T_{\mu\nu}(q) d^{PT}(q^2, \xi)+ \xi
L_{\mu\nu}(q) \Bigl] {1 \over q^2}
\end{equation}
remains undetermined within our approach. This was the price we
have had to pay to fix the functional dependence of the INP part
of the full gluon propagator (up to the arbitrary, in general,
residues). The only thing we know about the PT gluon form factor
$d^{PT}(q^2, \xi)$ is that it is a regular function at $q^2
\rightarrow 0$ and should satisfy AF at $q^2 \rightarrow \infty$.
Let us also note that it includes the free gluon propagator
$D^0_{\mu\nu}(q) = i[ T_{\mu\nu}(q) + \xi L_{\mu\nu}(q)] (1 /
q^2)$ as well.

We distinguish between the two terms in the full gluon propagator
(2.3) first by the explicit presence of the mass gap (when it
formally goes to zero then the only PT term survives). Secondly,
the INP part of the full gluon propagator is characterized by the
presence of severe power-type (or equivalently NP) infrared (IR)
singularities $(q^2)^{-2-k}, \ k=0,1,2,3,...$. So these IR
singularities are defined as more singular than the power-type IR
singularity of the free gluon propagator $(q^2)^{-1}$, which thus
can be defined as the PT IR singularity. Due to the character of
the IR singularity the longitudinal component of the full gluon
propagator should be included into its PT part, so its INP part
becomes automatically transversal.

Both terms in Eq. (2.3) are valid in the whole energy/momentum
range, i.e., they are not asymptotics. At the same time, we have
achieved the exact and unique separation between the two terms
responsible for the NP (dominating in the IR at $q^2 \rightarrow
0$) and the nontrivial PT (dominating in the ultraviolet (UV) at
$q^2 \rightarrow \infty$) dynamics in the true QCD vacuum. Thus it
is really beset with severe IR singularities. Within the general
nonlinear iteration solution they should be summarized
(accumulated) into the full gluon propagator and effectively
correctly described by its structure in the deep IR domain,
exactly represented by its INP part. Concluding, let us emphasize
that in performing the general nonlinear iteration procedure no
truncations/approximations and no special gauge choice have been
made in the corresponding regularized skeleton loop integrals,
which represent the different terms contributing to the full gluon
self-energy and hence to its subtracted counterpart.

\subsection{Subtraction(s)}

As emphasized in our previous works \cite{2,6}, many important
quantities in QCD, such as the gluon and quark condensates, the
topological susceptibility, the Bag constant, etc., are defined
only beyond the PT. This means that they are determined by such
$S$-matrix elements (correlation functions) from which all types
of the PT contributions should be, by definition, subtracted.
Anyway, to calculate correctly any truly NP quantity from first
principles in low-energy QCD one has to begin with making
subtractions at the fundamental quark-gluon level. Using the exact
decomposition (2.3), let us define the INP gluon propagator by the
corresponding subtraction as follows:

\begin{equation}
D^{INP}_{\mu\nu}(q; \Delta^2) = D_{\mu\nu}(q; \Delta^2)-
D_{\mu\nu}(q; \Delta^2=0) = D_{\mu\nu}(q; \Delta^2)-
D^{PT}_{\mu\nu}(q),
\end{equation}
so that the full gluon propagator becomes an exact sum of the two
different terms in complete agreement with Eq. (2.3). The
principal difference between the full gluon propagator
$D_{\mu\nu}(q; \Delta^2)$ and the INP gluon propagator
$D^{INP}_{\mu\nu}(q; \Delta^2)$ is that the latter one is free of
the PT contributions, while the former one, being also NP, is
"contaminated" by them. Also, the INP gluon propagator is
manifestly transversal, i.e., does not depend explicitly on the
gauge-fixing parameter. Since the formal PT limit $\Delta^2=0$ is
uniquely defined in the framework of our method, the separation
between the INP and PT gluon propagators is uniquely defined as
well. Evidently, the subtraction (2.7) is equivalent to the
subtraction made at the level of the full gluon form factor in Eq.
(2.1) as follows: $d(q^2) = d(q^2) - d^{PT}(q^2)+ d^{PT}(q^2)=
d^{INP} (q^2) + d^{PT}(q^2)$. It is worth emphasizing once more,
that making the above-defined subtraction, we are achieving the
two goals simultaneously: the transversality of the gluon
propagator relevant for the truly NP QCD, and it automatically
becomes free of the PT contributions ("PT contaminations") as
well. So our prescription for the subtraction at the fundamental
gluon level is simply reduced to the replacement of the general
iteration solution by its INP part everywhere, i.e.,

\begin{equation}
D_{\mu\nu}(q; \Delta^2) \longrightarrow D^{INP}_{\mu\nu}(q;
\Delta^2),
\end{equation}
and/or equivalently

\begin{equation}
d(q^2; \Delta^2) \longrightarrow d^{INP} (q^2; \Delta^2).
\end{equation}
Their explicit expressions are given below. The necessity of such
kind of the subtraction and other types ones has been discussed
and justified in our papers \cite{2,6} (see also references
therein), where some concrete examples are present as well. Let us
emphasize in advance that the replacements (2.8) and (2.9) for the
full gluon propagator and the similar one for the full ghost
self-energy (see below) mean omitting their corresponding PT parts
in which their corresponding free PT counterparts are to be
included.

 Concluding, the replacements (2.8) and/or (2.9) are
necessary to be made first at the fundamental gluon level in order
to correctly calculate from first principles any truly NP physical
quantities and processes in low-energy QCD.

\subsection{Multiplicative renormalizations}

Thus the full gluon propagator, which is relevant for the
description of the truly NP QCD dynamics, is as follows:

\begin{equation}
D_{\mu\nu}(q, \Delta^2) = i T_{\mu\nu}(q) {\Delta^2 \over (q^2)^2}
f(q^2),
\end{equation}
and

\begin{equation}
f(q^2) = \sum_{k=0}^{\infty} (\Delta^2 / q^2)^k \Phi_k(\lambda,
\alpha, \xi, g^2).
\end{equation}
Evidently, after making the above described subtraction (2.7) or
equivalently the replacement (2.8) the superscript "INP" has been
omitted in order to simplify notations.

A new surprising feature of this solution is that its both
asymptotics at zero ($q^2 \rightarrow 0$) and at infinity ($q^2
\rightarrow \infty$) are to be determined by its $(q^2)^{-2}$
structure only. This structure determines the behavior of the
solution (2.10) at infinity, since all other terms in this
expansion are suppressed in this limit. So the main problem with
our solution (2.10) is its structure in the deep IR region ($q^2
\rightarrow 0$). The function $f(q^2)$ is defined by its Laurent
expansion, and thus it has an isolated essentially singular point
at $q^2=0$. Its behavior in the neighborhood of this point is
regulated by the Weierstrass-Sokhocky-Kazorati (WSK) theorem
\cite{7} which tells that

\begin{equation}
\lim_{n \rightarrow \infty}f(q^2_n) = Z, \quad q^2_n \rightarrow
0,
\end{equation}
where $Z$ is an arbitrary number, and $\{q^2_n \}$ is a sequence
of points $q^2_1, q^2_2, ..., q^2_n$ along which $q^2$ goes to
zero, and for which this limit always exists. Of course, $Z$
remains arbitrary (it depends on the chosen sequence of points),
but in general it depends on the same set of parameters as the
residues, i.e., $Z \equiv Z(\lambda, \alpha, \xi, g^2)$. This
theorem thus allows one to replace the Laurent expansion $f(q^2)$
by $Z$ when $q^2 \rightarrow 0$ independently from all other test
functions in the corresponding integrands, i.e.,

\begin{equation}
f(0; \lambda, \alpha, \xi, g^2) \rightarrow Z(\lambda, \alpha,
\xi, g^2).
\end{equation}
There is no doubt that the only real severe (i.e., NP) IR
singularity of the full gluon propagator (2.10) is the
$(q^2)^{-2}$ NP IR singularity, while the Laurent expansion
$f(q^2)$ should be treated in accordance with the WSK theorem.

Our consideration at this stage is necessarily formal, since the
mass gap remains unrenormalized yet as well as all other
quantities. So far it has been only regularized, i.e., $\Delta^2
\equiv \Delta^2(\lambda, \alpha, \xi, g^2)$. However, due to the
above-formulated WSK theorem, the full gluon propagator (2.10)
effectively becomes

\begin{equation}
D_{\mu\nu}(q; \Delta^2) = i T_{\mu\nu}(q) { 1 \over (q^2)^2}
Z(\lambda, \alpha, \xi, g^2) \Delta^2 (\lambda, \alpha, \xi, g^2),
\end{equation}
so just its $(q^2)^{-2}$-structure is all that matters, indeed.
Before going to the $\lambda \rightarrow \infty$ limit in this
expression, let us note that in general the coupling constant
squared $g^2$ may also depend on $\lambda$, becoming thus the
so-called "running" effective charge $g^2 \sim \alpha_s(\lambda)$.
Let us now define the renormalized (R) mass gap in the strong
coupling regime as follows:

\begin{equation}
\Delta^2_R =  Z(\lambda, \alpha_s(\lambda)) \Delta^2 (\lambda,
\alpha_s(\lambda)), \quad \lambda \rightarrow \infty, \quad
\alpha_s(\lambda) \rightarrow \infty,
\end{equation}
at any arbitrary $\alpha$ and $\xi$, the explicit dependence on
which was omitted as unimportant. So that we consider $Z(\lambda,
\alpha_s(\lambda))$ as the multiplicative renormalization constant
for the mass gap, and $\Delta^2_R$ is the physical mass gap within
our approach. Precisely this quantity should be identified with
the Jaffe and Witten mass gap \cite{8} (due to the WSK theorem, we
can always choose such $Z$ in order to make $\Delta^2_R$ positive,
finite, gauge-independent, etc.). The two other possible types of
the effective charge's behavior when $\lambda \rightarrow \infty$
have been discussed in our previous work \cite{2}.

Thus the full gluon propagator relevant for the description of
truly NP QCD dynamics and expressed in terms of the renormalized
quantities finally becomes

\begin{equation}
D_{\mu\nu}(q; \Delta^2_R) = i T_{\mu\nu}(q) { \Delta^2_R \over
(q^2)^2}.
\end{equation}
The renormalization of the mass gap is an example of the NP
renormalization ( let us remind \cite{2} that an infinite number
of iterations (all iterations) invokes each severe IR singularity
labelled by $k$ in Eq. (2.11)). The corresponding initial
renormalization constant $Z(\lambda, \alpha, \xi, g^2)$ appears
naturally, so the general renormalizability of QCD is not
affected. Since we were able to accumulate all the quadratic
divergences (parameterized as the initial ("bare") mass gap) into
its renormalization, the $(q^2)^{-2}$-type behavior of the
relevant gluon propagator (2.16) at infinity is not dangerous any
more, i.e., it cannot undermine the general renormalizability of
QCD. It is worth reminding that in Ref. \cite{2} it has been
already explained why we call the potential (2.16) confining. In
our next papers we will show explicitly that it leads to the
confining quark propagator, indeed.

However, the real problem with our solution (2.16) is the behavior
at the origin ($q^2 \rightarrow 0$), since its IR singularity
represents the so-called severe IR singularity, and the PT fails
to deal with it. It should be treated by the distribution theory
(DT) \cite{9} into which the dimensional regularization method
(DRM) \cite{10} is to be correctly implemented (for a brief review
of this program see our previous work \cite{2} and references
therein). In order to show that our expression (2.16) is an exact
result, i.e., it is neither IR nor UV asymptotic, it is
instructive to begin with the initial expressions (2.10) and
(2.11), which are valid in the whole energy/momentum range.
Because of the summation over $k$, nothing should depend on it.
This is in agreement with what we already know from the WSK
theorem. Thus the only NP IR singularity of Eq. (2.10) is its
$(q^2)^{-2}$-structure. If $q$ is an independent skeleton loop
variable, then the dimensional regularization of this NP IR
singularity is given by the expansion \cite{2,9}

\begin{equation}
(q^2)^{- 2} = { 1 \over \epsilon} \Bigr[ \pi^2 \delta^4(q) +
O(\epsilon) \Bigl], \quad \epsilon \rightarrow 0^+.
\end{equation}
Here and below $\epsilon$ is the IR regularization parameter
(which determines the deviation of the number of dimensions from
four \cite{2,9,10}). It should go to zero at the final stage only.
Due to the $\delta^4(q)$ function in the residue of this
expansion, all the test functions which appear under corresponding
skeleton loop integrals should be finally replaced by their
expression at $q=0$. So Eq. (2.10) effectively becomes

\begin{equation}
D_{\mu\nu}(q; \Delta^2_R) = { 1 \over \epsilon}i T_{\mu\nu}(q)
\Delta^2 f(0) \delta^4(q)= { 1 \over \epsilon}i T_{\mu\nu}(q)
\Delta^2_R \delta^4(q),
\end{equation}
where the the replacement (2.13) (i.e., the result of the WSK
theorem) and the definition (2.15) have been used (the finite
number $\pi^2$ as usual is included into the renormalized mass
gap). For simplicity, the terms of the order $O(\epsilon)$ are not
shown. Evidently, substituting the expansion (2.17) into Eq.
(2.16), one obtains the same Eq. (2.18). This clearly shows that
the previous Eq. (2.16) is exact, i.e., it is not IR asymptotic,
and thus remain valid in the whole energy/momentum range.

The only problem remaining to solve is how to remove the pole $1/
\epsilon$ which necessarily appears in the full gluon propagator.
As emphasized in Ref. \cite{2}, in the presence of severe IR
singularities, which are to be regularized in terms of the IR
regularization parameter $\epsilon$ via the expansion (2.17), in
general, all the Green's functions and parameters depend on it.
The only way to remove the pole in $\epsilon$ from the full gluon
propagator (2.18) is to define the IR renormalized mass gap as
follows:

\begin{equation}
\Delta^2_R = X(\epsilon) \bar{\Delta}^2_R = \epsilon
\bar{\Delta}^2_R, \quad \epsilon \rightarrow 0^+,
\end{equation}
where $X(\epsilon) = \epsilon$ is the IR multiplicative
renormalization (IRMR) constant for the mass gap, and the IR
renormalized mass gap $\bar{\Delta}^2_R$ exists as $\epsilon
\rightarrow 0^+$, by definition, contrary to $\Delta^2_R$. In both
expressions for the mass gap the dependence on $\epsilon$ is
assumed but not shown explicitly. Thus the IR and UV renormalized
gluon propagator becomes

\begin{equation}
D_{\mu\nu}(q; \bar{\Delta}^2_R) = i T_{\mu\nu}(q) \bar{\Delta}^2_R
\delta^4(q),
\end{equation}
and it is instructive to compare it with the initial solution
(2.10), which was neither UV nor IR renormalized. It has been only
regularized. However, it survived both renormalization programs.
In this paper we will show that the IR renormalizaion of the full
gluon propagator or equivalently of the mass gap is completely
sufficient to remove all severe IR singularities from all the
skeleton loop integrals which may appear in the INP QCD. However,
let us note in advance that beyond the one-loop skeleton integrals
the analysis should be done in a more sophisticated way, otherwise
the appearance of the product of at least two $\delta$ functions
at the same point is possible. However, this product is not
defined in the DT \cite{9}. So in the multi-loop skeleton diagrams
instead of the $\delta$ functions in the residues their
derivatives may appear \cite{2,9}. They should be treated in the
sense of the DT.

Concluding, {\bf Eq. (2.16) is an exact result, i.e., it is
neither UV nor IR asymptotic, manifestly transversal and even
implicitly does not depend on the gauge-fixing parameter. If $q$
is an independent skeleton loop variable, then Eq. (2.20) is to be
used from the very beginning}.

\subsection{The ZMME quantum structure of the true QCD ground state}

The true QCD ground state is in principle a very complicated
confining medium, containing many types of gluon field
configurations, components, ingredients and objects of different
nature \cite{1,3,4,11,12}. Its dynamical and topological
complexity means that its structure can be organized at both the
quantum and classical levels. It is definitely "contaminated" by
such gluon field excitations and fluctuations, which are of the PT
origin, nature and magnitude. Moreover, it may contain such extra
gluon field configurations, which cannot be described as possible
solutions to the QCD dynamical equations of motion, either quantum
or classical, for example, the vortex-type ones \cite{13}. The
only well known classical component of the QCD ground state is the
topologically nontrivial instanton-antiinstanton type of
fluctuations of gluon fields, which are solutions to the Euclidean
Yang-Mills (YM) classical equations of motion in the weak coupling
regime \cite{14,15}. However, they are by no means dominant but,
nevertheless, they play a special role in the QCD vacuum. In our
opinion their main task is to prevent quarks and gluons to freely
propagate in the QCD vacuum. It seems to us that this role does
not contradict their standard interpretation as tunneling
trajectories linking vacua with different topology (\cite{1,15}
and references therein).

Our quantum-dynamical approach to the true QCD ground state is
based on the existence and the importance of such kind of the NP
excitations and fluctuations of virtual gluon fields which are
mainly due to the NL interactions between massless gluon modes
without explicitly involving some extra degrees of freedom. It
analytically takes into account such gluon field configurations
which can be described by the general nonlinear iteration solution
(in the form of the corresponding skeleton loops expansion) to the
QCD quantum equation of motion for the full gluon propagator in
the presence of a mass gap. This solution inevitably becomes
plagued by severe IR singularities, which thus play an important
role in the large-distances behavior of QCD. They are to be
summarized (accumulated) into the purely transversal part of the
full gluon propagator, and are to be effectively correctly
described by its severely singular structure in the deep IR
domain, Eq. (2.10). We will call them the purely transversal
singular gluon fields. In other words, they represent the purely
transversal quantum virtual fields with the enhanced low-frequency
components/large scale amplitudes due to the NL dynamics of the
massless gluon modes.

At this stage it is difficult to identify actually which type of
gauge field configurations can be finally formed by the purely
transversal singular gluon fields in the QCD ground state, i.e.,
to identify relevant field configurations: chromomagnetic,
self-dual, stochastic, etc. However, if these gauge field
configurations can be absorbed into the gluon propagator (i.e., if
they can be considered as solutions to the corresponding SD
equation), then its severe IR singular behavior is a common
feature for all of them. Being thus a general phenomenon, the
existence and the importance of quantum excitations and
fluctuations of severely singular IR degrees of freedom inevitably
lead to the general zero momentum modes enhancement (ZMME) effect
in the QCD ground state (or equivalently ZME which means simply
zero momentum enhancement). Thus our approach to the true QCD
ground state, based on the general ZMME phenomenon there, can be
analytically formulated in terms of the full gluon propagator
(2.10). Moreover, it has been clearly shown that our solution
survives both renormalization programs, and is explicitly given in
Eq. (2.19). At the same time, the above-mentioned possible
complications due to the multi-loop skeleton diagrams should be
always kept in mind.

Working always in the momentum space, we are speaking about the
purely transversal singular gluon fields responsible for color
confinement in our approach. Discussing the relevant field
configurations, we always will mean the functional space. Speaking
about relevant field configurations (chromomagnetic, self-dual,
stochastic, etc), we mean all the low-frequency modes of these
virtual transversal fields. Only large scale amplitudes of these
fields ("large transversal gluon fields") are to be taken into
account by the INP part of the full gluon propagators. All other
frequencies are to be taken into account by corresponding PT part
of the gluon propagators. Apparently, it is not correct to speak
about specific field configurations that are solely responsible
for color confinement. The low-frequency components/large scale
amplitudes of all the possible in the QCD vacuum the purely
transversal virtual fields are important for the dynamical and
topological formation of such gluon field configurations which are
responsible for color confinement and other NP effects within our
approach to low-energy QCD. For convenience, we will call them the
purely transversal severely singular gluon field configurations as
mentioned above.

The ZMME (or simply ZME) mechanism of quark confinement is nothing
but the well forgotten IR slavery (IRS) one, which can be
equivalently referred to as a strong coupling regime \cite{1,16}.
Indeed, at the very beginning of QCD the general idea
\cite{16,17,18,19,20,21,22,23} was expressed that because of the
self-interaction of massless gluons in the QCD vacuum, the quantum
excitations of the IR degrees of freedom enable us to understand
confinement, dynamical (spontaneous) breakdown of chiral symmetry
and other NP effects. In other words, the importance of the deep
IR structure of the true QCD vacuum has been emphasized as well as
its relevance to the above-mentioned NP effects and the other way
around. This development was stopped by the wide-spread wrong
opinion that severe IR singularities cannot be put under control.
Here we have explicitly shown (see also our recent papers
\cite{2,24,25} and references therein) that the adequate
mathematical theory of quantum YM physical theory is the DT (the
theory of generalized functions) \cite{9}, complemented by the DRM
\cite{10}. Together with the theory of functions of complex
variable \cite{7} they provide a correct treatment of these severe
IR singularities without any problems. Thus, we come back to the
old idea but on a new basis that is why it becomes new ("new is
well forgotten old"). In other words, we put the IRS mechanism of
quark confinement on a firm mathematical ground.

Concluding, there is no doubt that the purely transversal severely
singular virtual gluon field configurations play an important role
in the dynamical and topological structure of the true QCD ground
state, leading thus to the general ZMME effect there. The quark,
ghost Green's functions and the corresponding ST identities, etc.
should be then reconstructed on the basis of this effect. This
makes it possible to take into account the response of the NP QCD
vacuum.

\section{Quark sector}

Together with the full gluon propagator, the full quark propagator
also plays one of the most important roles in QCD. After
establishing the confining gluon propagator in the previous
section, the next step is to derive the confining quark
propagator. It allows one to make further necessary steps in the
realization of the program to calculate physical
observables/processes in low-energy QCD from first principles. The
quark Green's function satisfies its own SD equation with
Euclidean signature, namely

\begin{equation}
S^{-1} (p) = S_0^{-1} (p)- C_F \int {i d^4q \over (2 \pi)^4}
\Gamma_{\mu} (p,q) S(p-q) \gamma_{\nu} D_{\mu\nu}(q),
\end{equation}
and $C_F$ is the eigenvalue of the quadratic Casimir operator in
the fundamental representation (for $SU(N_c)$, in general, $C_F=
(N_c^2-1)/2N_c= 4/3$ at $N_c=3$). $\Gamma_{\mu} (p,q)$ is the
quark-gluon proper vertex, while $S(p)$ is the full quark
propagator. Here and everywhere below the dependence on the
coupling constant in the corresponding powers which comes from the
corresponding point-like vertices has been included in the
corresponding proper vertices. Let us remind that in the presence
of the mass gap it plays no any role, anyway. The free quark
propagator is

\begin{equation}
S_0^{-1}(p)= i( \hat p + m_0)
\end{equation}
with $m_0$ being the current ("bare") quark mass.

Since $q$ is the independent skeleton loop variable and the number
of skeleton loops coincides with the number of the full gluon
propagators, we can directly substitute our solution for the
confining gluon propagator (2.20), which yields

\begin{equation}
S^{-1} (p) = S_0^{-1} (p)+ \bar{\Delta}^2_R \Gamma_{\mu} (p,0)
S(p) \gamma_{\mu},
\end{equation}
and, for convenience, all other finite numerical factors have been
included into the mass gap with retaining the same notation. In
deriving this equation, we have used the confining gluon
propagator which was already UV and IR renormalized, i.e., free
from all types of UV divergences and IR singularities,
parameterized in terms of $\epsilon$. In other words, the quark SD
equation (3.3) is free from all these problems. So, we can
consider all other Green's functions entering this equation,
namely the quark-gluon proper vertex and th full quark propagator,
as the UV and IR renormalized from the very beginning, and
omitting the corresponding subscripts and bars, for simplicity. In
what follows we will always replace any Green's functions by their
IR renormalized counterparts when there will be no explicit
dependence on $\epsilon$ like it was in this case.

However, one important issue should be discussed in more detail in
advance. In passing from Eq. (3.1) to Eq. (3.3) it was implicitly
assumed that the vertex function $\Gamma_{\mu} (p,0)$ can be
simply obtained from $\Gamma_{\mu} (p,q)$ in the $q \rightarrow 0$
limit. Evidently, this is only possible if the vertex is a regular
function of the momentum transfer $q$. In principle, we did not
specify the analytical properties of all the vertex functions with
respect to their gluon momenta transfer when the confining gluon
propagator has been derived in Ref. \cite{2}. At the level of the
gluon SD equation and within its nonlinear iteration solution the
analytical properties of the vertex functions were not crucial.
However, beyond the gluon sector they may be important. For
example, if the proper vertex in Eq. (3.1) has additional
singularities with respect to the gluon momentum $q$, then they
can be effectively incorporated into the gluon propagator itself.
The initial singular structure $(q^2)^{-2}$ of Eq. (2.10) becomes
more complicated, so instead of the exponent $-2$ a more general
exponent $-2-k, \ k=0,1,2,3...$ will appear (and there is no
summation over $k$, i.e., each $k$ is to be investigated
independently). At the same time, the different $k$ mean different
solutions, and different solutions mean different vacua (see
discussion in Appendix B of Ref. \cite{2}). The Lagrangian of QCD
formally remains the same, while the theory is completely
different from the normal QCD. By it we mean QCD in which the zero
momenta transfer limit exists in all QCD vertex functions. Anyway,
the normal QCD (which obviously corresponds to $k=0$) should be
investigated independently from QCD with additional singularities
in the vertex functions. In what follows it is assumed that all
severe IR singularities can be summarized by the full gluon
propagator, and thus all the vertex functions are regular
functions of the corresponding momenta transfer. Let us emphasize
once more that this is obviously not a restriction, moreover
important it may be the most realistic case. It is worth noting as
well in advance that the smoothness properties of the
corresponding test functions (which will be established in the
subsequent paper) are in complete agreement with the
above-mentioned regularity of all the QCD vertices.

\section{Ghost sector}

The information about the quark-gluon vertex function at zero
momentum transfer, needed for the evaluation of the confining
quark propagator (3.3), can be provided by the quark ST identity
\cite{1,17,26,27} (and references therein), which contains unknown
ghost contributions in the covariant gauge. For this reason let us
consider in this section the SD equation for the ghost self-energy
$b(k^2)$, which also obeys a simple SD equation with Euclidean
signature \cite{17,27}

\begin{equation}
i k^2 b(k^2) = - C_A \int {i d^4q \over {(2\pi)^4}} G_\mu(k,q)
G(k-q) (k-q)_\nu D_{\mu\nu}(q),
\end{equation}
where $C_A$ is the eigenvalue of the quadratic Casimir operator in
the adjoint representation (for $SU(N_c)$, in general $C_A = N_c =
3$). The full ghost propagator is

\begin{equation}
G(k) = - {i \over {k^2\left[ 1 + b(k^2) \right]}}
\end{equation}
and

\begin{equation}
G_\mu(k,q) = k^\lambda G_{\lambda\mu}(k,q)
\end{equation}
is the ghost-gluon proper vertex ( $G_{\lambda\mu} =
g_{\lambda\mu}$ in the PT).

As for the quark SD equation, in the gluon self-energy the
momentum transfer $q$ is the independent skeleton loop variable.
This allows one to directly substitute again Eq. (2.20) which
yields

\begin{equation}
ik^2 b^{INP}(k^2) = \bar{\Delta}^2_R G_{\mu}(k,0) G(k)k_{\mu},
\end{equation}
where again all finite numerical factors have been included into
the mass gap. We also retain the superscript "INP" for the gluon
self-energy in the left-hand-side of this equation for future
purpose (see next section). This is instructive to do in order to
indicate that its right-hand-side has been obtained by replacing
the full gluon propagator by its INP counterpart in accordance
with our method.

It is convenient to rewrite Eq. (4.4) in the equivalent form as
follows:

\begin{equation}
- \bar{\Delta}^2_R  G_\mu(k,0) G(k) = i k_{\mu} b^{INP} (k^2).
\end{equation}
Just this equation will be used in order to investigate the
quark-gluon vertex function at zero momentum transfer. In the
corresponding ST identity the momentum transfer goes through the
ghost momentum (see next section). For that very reason, let us
assume that the ghost self-energy $b(k^2)$ exists and is finite at
$k^2=0$. Evidently, this means that both terms, namely
$b^{INT}(k^2)$ and $b^{PT}(k^2)$, which appear in the formal
decomposition $b(k^2)=b(k^2) - b^{PT}(k^2)+ b^{PT}(k^2)=
b^{INT}(k^2)+ b^{PT}(k^2)$, also exist and are finite at zero
point (in agreement with the above-mentioned regularity of the QCD
vertex functions with respect to their momenta transfer). This can
be directly shown, but we will not complicate the context of this
section, since our final results will not depend explicitly on
this auxiliary technical assumption. Concluding, let us only note
that the above-mentioned decomposition follows from the exact and
unique subtraction (2.7) after its substitution into the ghost
self-energy SD equation (4.1).

\section{Quark-ghost sector}

Though nothing should explicitly depend on ghost degrees of
freedom in QCD, nevertheless, the ghost-quark sector contains a
very important piece of information on quark degrees of freedom
themselves through the corresponding quark ST identity. Precisely
this information should be self-consistency taken into account.
Otherwise any solutions to the dynamical equations will be plagued
by unusual analytical properties (unphysical singularities), since
in the absence of ghosts the unitarity of $S$-matrix is violated.
The ST identity for the quark-gluon vertex function
$\Gamma_\mu(p,k)$ is (\cite{1,17,26,27,28,29,30} and references
therein)

\begin{equation}
- i k_\mu \Gamma^a_\mu(p,k) \left[ 1 + b(k^2) \right] = \left[ T^a
- B^a(p,k)\right] S^{-1}(p+k) -  S^{-1}(p)\left[T^a -
B^a(p,k)\right],
\end{equation}
where $b(k^2)$ is the ghost full self-energy and $B^a(p,k)$ is the
ghost-quark scattering amplitude. $T^a$ are the color group
generators. From the ST identity (5.1) one recovers the standard
QED-type Ward-Takahashi (WT) identity in the formal $b = B = 0$
limit.

The ghost-quark scattering kernel $B^a(p,k)$ is determined by its
skeleton expansion

\begin{equation}
B^a (p, k) = \sum_{n=1}^{\infty} B^a_n (p, k),
\end{equation}
which diagrammatical representation can be found, for example in
Refs. \cite{17,27,29,30}. In the Landau gauge ($\xi =0$) and at
$k=0$ Taylor \cite{28} has shown that it is zero, i.e.,

\begin{equation}
B^a (p, 0) = 0,
\end{equation}
and this is valid for each skeleton term in the skeleton expansion
(5.2), i.e., the relation (5.3) is valid because each $B^a_n (p,
0)=0$ in the Landau gauge.

Let us begin with the investigation of the first term $B_1(p,k)$
of the $B(p,k)$ skeleton expansion (5.2). After the evaluation of
the color group factors its analytical expression becomes
(Euclidean space)

\begin{equation}
B_1(p,k) = - {1\over 2} C_A \int {i d^4 q \over {(2\pi)^4}} S(p-q)
\Gamma_\nu(p-q, q) G_\mu(k, q) G(k+q) D_{\mu\nu}(q).
\end{equation}
Before proceeding further, let us show explicitly that it
satisfies the Taylor's general relation (5.3). In the Landau gauge
$D_{\mu\nu}(q) \sim T_{\mu\nu}(q)$ and at $k=0$ the ghost-gluon
vertex $G_\mu(0, q) \sim q_{\mu}$, so $q_{\mu} T_{\mu\nu}(q)=0$
leads to $B_1(p,0) =0$, indeed, in the Landau gauge. These
arguments are valid term by term in the skeleton expansion (5.2).

As in previous cases the gluon momentum $q$ is independent
skeleton loop variable, so again Eq. (2.20) can be directly
substituted, which yields

\begin{equation}
B_1(p, k) =  {1\over 2} \bar{\Delta}^2_R S(p) \Gamma_\mu(p,0)
G_\mu(k,0) G(k),
\end{equation}
and using further the ghost SD equation (4.5), one finally obtains

\begin{equation}
B_1(p, k) = - {1\over 2}i S(p) \Gamma_\mu(p,0) b^{INP} (k^2)
k_\mu,
\end{equation}
which clearly shows that it is of order $k$ ($\sim O(k)$) when $k$
goes to zero since $b^{INP}(0)$ exists and finite (see previous
section). Let us emphasize that this final expression does not
depend on the mass gap as it should be. Moreover, in the
expression (5.5) the mass gap (after the inclusion of all finite
numerical factors) is the same as in Eq. (4.5), since the ghost
line in the expression (5.4) is the same as in Eq. (4.1).

 The analytical expression of the second skeleton diagram for the
 ghost-quark scattering kernel $B(p,k)$ is

\begin{eqnarray}
&B_2(p,k)& = A  \int {id^4 q \over {(2\pi)^4}}  \int {id^4 l \over
{(2\pi)^4}} S(p-q+l) \Gamma_{\beta}(p-q+l, l) S(p-q)
\nonumber\\
&\Gamma_\nu(p-q, q)& G_\mu(k, -l) G(k-l)G_{\alpha}(k-l,q) G(k-l+q)
D_{\mu\beta}(l) D_{\alpha\nu}(q),
\end{eqnarray}
where the constant A is a result of the summation over color group
indices (its explicit expression is not important here and below).
Since both gluon momenta $q$ and $l$ are independent skeleton loop
variables, we again can use Eq. (2.20) twice, which yields

\begin{equation}
B_2(p,k) = A_0 \bar{\Delta}_R^4 S(p) \Gamma_{\beta}(p, 0) S(p)
\Gamma_\nu(p, 0) G_\beta(k, 0) G(k) G_{\nu}(k,0) G(k),
\end{equation}
and again using the ghost SD equation (4.5) twice, one finally
obtains

\begin{equation}
B_2(p,k) = \bar A_0 S(p) \Gamma_{\beta}(p, 0) S(p) \Gamma_\nu(p,
0) [b^{INP} (k^2)]^2 k_\beta k_{\nu},
\end{equation}
which clearly shows that this term is of order $k^2$ as it goes to
zero, since $ b^{INP} (k^2)$ is finite at zero point.

In the same way it is possible to show that the third term
$B_3(p,k)$ is of the order $k^3$ as $k$ goes to zero. These
arguments are valid term by term in the skeleton expansion for the
ghost-quark scattering kernel $B(p,k)$ (5.2). So, we have an exact
estimate

\begin{equation}
B_n(p,k) = O(k^n), \qquad k \rightarrow 0.
\end{equation}
It means that we maintain Taylor's general result (5.3). It is
worth emphasizing, however, that our confining gluon propagator is
automatically transversal, i.e., we did not choose the Landau
gauge by hand.

Differentiating now the quark ST identity (5.1) with respect to
$k_\mu$ and passing to the limit $k=0$, one obtains ($d_{\mu} = d
/ dp_{\mu}$, by definition)

\begin{equation}
- i \Gamma_\mu(p, 0) \left[ 1 + b(0) \right] = d_\mu S^{-1} (p) -
\Psi_\mu (p) S^{-1}(p) + S^{-1}(p)\Psi_\mu (p),
\end{equation}
where $\Psi_\mu (p)$ is defined as

\begin{equation}
\Psi_\mu (p) = \left[ { \partial \over \partial k_\mu } B (p, k)
\right]_{k=0}  = - {1\over 2}i b^{INP} (0) S(p) \Gamma_\mu(p,0),
\end{equation}
since due to an estimate (5.10) the first term (5.6) survives only
in the $k=0$ limit. Substituting it back into the ST identity
(5.11), one obtains that it becomes

\begin{equation}
\left[ 1 +  b^{PT}(0) + {1 \over 2} b^{INP}(0) \right]
\Gamma_\mu(p,0) = i d_\mu S^{-1}(p) -{1 \over 2} b^{INP}(0) S(p)
\Gamma_\mu(p,0) S^{-1}(p),
\end{equation}
where the above-mentioned formal decomposition $b(0) = b^{PT}(0) +
b^{INP}(0)$ has been also used (let us recall, however, that this
decomposition is exact and unique, since it is due to the
substitution of the subtraction (2.7) into the ghost self-energy
SD equation (4.1)). In this form the quark ST identity first has
been obtained by Pagels in his pioneering paper on NP QCD
\cite{17}. However, this form is not acceptable, since it depends
explicitly on the PT part of the ghost self-energy, i.e., it is
not completely free yet from the PT contributions
("contaminations").

Fortunately, we already know how to solve this problem. In
accordance with our subtraction prescription (2.8) the full ghost
self-energy at zero $b(0)$ should be replaced by its INP part
$b^{INP}(0)$, which is equivalent to omit in the quark ST identity
(5.13) the PT part of the ghost self-energy in which its free PT
counterpart $b^{PT}_0=1$ is to be included. In other words, the
sum $1 + b^{PT}(0)= b^{PT}_0 + b^{PT}(0) \rightarrow b^{PT}(0)$
should be omitted in the left-hand-side of the quark SD identity
(5.13). So one gets

\begin{equation}
 {1 \over 2} b^{INP}(0) \Gamma_\mu(p,0) = i d_\mu
S^{-1}(p) -{1 \over 2} b^{INP}(0) S(p) \Gamma_\mu(p,0) S^{-1}(p),
\end{equation}
and thus it becomes free of all types of the PT contributions,
indeed. At the same, the necessary information on quark degrees of
freedom important for the INP QCD dynamics has been completely
extracted from the initial ST identity (the second term in Eq.
(5.14), while the first term is the standard WT-type one). In a
more sophisticated way this procedure is described in Appendix A.

\section{ Intrinsically Nonperturbative (INP) QCD}

Let us now write down the system of equations obtained in the
quark sector

\begin{eqnarray}
S^{-1} (p)&=& S_0^{-1} (p)+ \bar{\Delta}_R^2 \Gamma_\mu(p,0) S(p)
\gamma_\mu,
\nonumber\\
{1 \over 2} b^{INP}(0) \Gamma_\mu(p,0) &=& id_\mu S^{-1}(p) - {1
\over 2} b^{INP}(0) S(p) \Gamma_\mu(p,0) S^{-1}(p).
\end{eqnarray}
This system still suffers from the explicit presence of the
unknown number, namely $b^{INP}(0)$. To resolve this difficulty,
let us rescale the proper vertex as follows:

\begin{equation}
{1 \over 2} b^{INP}(0) \Gamma_\mu(p,0) \Longrightarrow
\Gamma_\mu(p,0),
\end{equation}
which makes it possible to include this unknown number into the
final mass gap, which we denote as $\Lambda_{NP}^2$. The initial
system of equations (6.1) then becomes

\begin{eqnarray}
S^{-1} (p)&=& S_0^{-1} (p)+ \Lambda_{NP}^2 \Gamma_\mu(p,0) S(p)
\gamma_\mu,
\nonumber\\
\Gamma_\mu(p,0) &=& id_\mu S^{-1}(p) - S(p) \Gamma_\mu(p,0)
S^{-1}(p).
\end{eqnarray}
Let us emphasize once more that the obtained system of equations
(6.3) is exact, i.e., no approximations/truncations have been made
so far. Formally it is valid in the whole energy/momentum range,
but depends only on the mass gap responsible for the true NP QCD
dynamics. It is free from all the types of the PT contributions
("PT contaminations") at the fundamental quark-gluon level. Also,
it is manifestly gauge-invariant, i.e., does not depend explicitly
on the gauge-fixing parameter. In the part II of this paper it
will be our primary goal to solve this system. For the first time
the system of equations (6.3) has been published in our
preliminary papers \cite{29,30}.

We consider the INP QCD as a true theory of low-energy QCD, which
makes it possible to calculate the physical observables/processes
in QCD from first principles. Let us recall that we define INP QCD
(see Refs. \cite{2,6} and section II in this work) by the
subtractions of all the types and at all levels of the PT
contributions from the corresponding QCD expressions, equations,
relations, etc. Symbolically this can be shown as follows:

\begin{equation}
QCD \Longrightarrow INP \ QCD = QCD - GPT \ QCD,
\end{equation}
where, evidently, $GPT \ QCD$ symbolically stands for the general
PT (GPT) QCD, and which includes all of the mentioned PT
contributions. The first necessary subtraction has been done at
the fundamental gluon level in Eq. (2.7). All other related
subtractions have been also made in the quark, ghost and quark ST
identity sectors in order to get to the final system of equations
(6.3) at the fundamental quark-gluon level. It allows one to
derive the full quark propagator in closed form and then to apply
such a quark propagator for the calculation of any physical
observable/process from first principle in terms of the mass gap
$\Lambda_{NP}^2$ in low-energy QCD.

Before going to some conclusions, it is worth making a few
remarks. Contrary to ghost and gluon degrees of freedom in which
their free PT counterparts have been included into the their
nontrivial PT parts, the free PT quark propagator has not been
subtracted in Eq. (6.3). Evidently, it has to be retained in order
to maintain the chiral limit physics in QCD, which is important to
correctly understand the structure of QCD at low energies.

Concluding, using the confining gluon propagator the corresponding
system of equations in the quark sector (6.3) has been derived in
a self-consistent way. It is free from all types of the PT
contributions, and thus is UV finite (i.e., free from the UV
divergences). It does not depend explicitly on the gauge-fixing
parameter. It has been derived for the Green's functions which
have been treated as the IR renormalized from the very beginning,
since the confining gluon propagator used (2.20) was the UV and IR
renormalized as well. However, the nontrivial IR renormalization
program can be performed. In this way one obtains the system of
the IR convergence conditions for the corresponding IRMR
constants, which relate the regularized quantities to their
renormalized counterparts. This makes it possible to remove all
severe IR singularities parameterized in terms of the IR
regularization parameter $\epsilon$ from all the equations,
identities, etc. in a self-consistent way. Its solution will lead
finally to the same system of equations (6.3), of course. That is
why there is no need in these technical complications if it is not
really necessary. This necessity may only appear in the multi-loop
skeleton diagrams, containing the three- and four-gluon proper
vertices.

\begin{acknowledgments}

Support in part by HAS-JINR Scientific Collaboration Fund and
Hungarian OTKA-T043455 grant (P. Levai) is to be acknowledged. I
would like to thank J. Nyiri for useful remarks, constant support
and help.

\end{acknowledgments}

\appendix

\section{Rescaling procedure}

Let us formulate the rescaling procedure in a more sophisticated
and general ways. Its final goal is to get the system of
equations, consisting of the quark SD equation (3.3) and quark ST
identity (5.13), free of all types of the PT contributions ("PT
contaminations"). For this purpose, it makes sense to rescale the
vertex in Eq. (5.13) as follows:

\begin{equation}
\left[ 1 + b^{PT}(0) + {1 \over 2} b^{INP}(0) \right]
\Gamma_\mu(p,0) \Longrightarrow \Gamma_\mu(p,0).
\end{equation}
The ST identity (5.13) then becomes

\begin{equation}
\Gamma_\mu(p,0) = id_\mu S^{-1}(p) - (1 + \delta)^{-1} S(p)
\Gamma_\mu(p,0) S^{-1}(p),
\end{equation}
where

\begin{equation}
\delta = {2 [1 + b^{PT}(0)] \over b^{INP}(0)}.
\end{equation}
Let us note here that $b^{PT}_0 =1$, where $b^{PT}_0$ is the form
factor of the free ghost propagator (see Eq. (4.2) at $b(k^2)=0$).
As in the case of the PT part of the full gluon propagator, let us
include it into the PT part of the full gluon self-energy with
retaining the same notation, i.e., the replacement $1 + b^{PT}(0)=
b^{PT}_0 + b^{PT}(0) \rightarrow b^{PT}(0)$ is understood in what
follows.

Expanding formally in powers of $\delta$, one gets

\begin{equation}
(1 + \delta)^{-1} = 1 + \sum_{n=2}^{\infty} (-1)^{n-1}
\delta^{n-1}, \qquad \delta = 2 {b^{PT}(0) \over b^{INP}(0)}.
\end{equation}
Substituting this back into the ST identity (A2), one obtains

\begin{equation}
\Gamma_\mu(p,0) = id_\mu S^{-1}(p) - S(p) \Gamma_\mu(p,0)
S^{-1}(p) + \Gamma'_{\mu}(p, 0),
\end{equation}
where, obviously,

\begin{equation}
\Gamma'_{\mu} (p, 0) = - S(p) \Gamma_\mu(p,0) S^{-1}(p)
\sum_{n=2}^{\infty} (-1)^{n-1} \Big( 2{b^{PT}(0) \over b^{INP}(0)}
\Big)^{n-1}.
\end{equation}

Making the same rescaling trick (A1) in the quark SD equation
(3.3), one obtains

\begin{equation}
S^{-1} (p)= S_0^{-1} (p)+ \Lambda_{NP}^2 \Gamma_\mu(p,0) S(p)
\gamma_\mu +  i \Sigma' (p),
\end{equation}
where, evidently,

\begin{equation}
i \Sigma' (p) = \Lambda_{NP}^2 \Gamma_\mu(p,0) S(p) \gamma_\mu
\sum_{n=2}^{\infty} (-1)^{n-1} \Big( 2 { b^{PT}(0) \over
b^{INP}(0)} \Big)^{n-1}.
\end{equation}
We also include the finite numerical factor $(2 / b^{INP}(0))$
into the mass gap $\bar{\Delta}^2_R$ and denote it as
$\Lambda_{NP}^2$. Just this quantity will be treated in what
follows as the physical mass gap responsible for the truly NP QCD
dynamics within our approach. Thus with the help of the formulated
rescaling procedure we were able to exactly identify (decouple)
the terms which are "contaminated" by the PT contributions due to
ghosts in the quark ST identity and the quark SD equation in Eqs.
(A5) and (A7), respectively. At the same, the necessary
information on quark degrees of freedom important for the INP QCD
dynamics has been completely extracted from the initial ST
identity (the second term in Eq. (A5), while the first term is the
standard WT-type one).

Let us now write down the system of equations obtained in the
quark sector

\begin{eqnarray}
S^{-1} (p)&=& S_0^{-1} (p)+ \Lambda_{NP}^2 \Gamma_\mu(p,0) S(p)
\gamma_\mu +  i \Sigma' (p),
\nonumber\\
\Gamma_\mu(p,0) &=& id_\mu S^{-1}(p) - S(p) \Gamma_\mu(p,0)
S^{-1}(p) + \Gamma'_{\mu} (p, 0),
\end{eqnarray}
and where the terms which are "contaminated" by the PT
contributions due to ghosts $i \Sigma' (p)$ and $\Gamma'_{\mu} (p,
0)$ are shown explicitly in Eqs. (A8) and (A6), respectively. The
first of this system of equations, namely the quark SD one depends
explicitly on the mass gap $\Lambda_{NP}^2$, which determines the
large-scale structure of the true QCD vacuum. In deriving this
system no approximation/truncations have been made by hand.
However, the two serious problems still remain to solve. The first
one is its above-mentioned "contamination" with the PT
contributions due to ghosts. The second one is in close relation
with the first one, namely the PT part of the ghost self-energy
$b^{PT}(0)$ may still depend explicitly on the gauge-fixing
parameter. Let us recall that we define INP QCD (see symbolic Eq.
(6.4)) by the subtractions of all the types and at all levels of
the PT contributions from the corresponding QCD expressions,
equations, relations, etc. Equivalently, this can be achieved by
simply dropping the terms "contaminated" by the PT contributions
in all equations, relations, etc. Doing so in the system of
equations (A9), one finally arrives to the same system of
equations (6.3) as it should be.

\end{document}